\documentclass[AMA,STIX1COL]{WileyNJD-v2}

\articletype{Research Article}%

\received{Will be added}
\revised{Will be added}
\accepted{Will be added}

\usepackage{lineno,hyperref}
\usepackage{setspace}
\usepackage{xcolor} 	
\usepackage{tabularx}
\usepackage{multirow}
\usepackage{subfig}
\usepackage{csvsimple}
\usepackage{caption}
\usepackage{array,multirow}
\usepackage{todonotes}
\usepackage{pgfplots}
\usepackage{anyfontsize}
\usepackage{tabto}

\newcommand{\vekt}[1]{\mathbf{#1}}
\newcommand{\vel}[0]{\vekt{u}_f}

\newcommand{\xtil}[0]{\vekt{\tilde{x}}}
\newcommand{\Jac}[0] { \widehat{\vekt{J}^{-1}}}
\newcommand{\JacN}[1] { \mbox{}^{#1} \Jac}
\newcommand{\CauchyStress}[0]{\vekt{T}}
\newcommand{\nsd}[0]{{n_{sd}}}
\newcommand{\fluidSolver}[0]{{\boldsymbol{\mathcal{F}}}}
\newcommand{\structSolver}[0]{{\boldsymbol{\mathcal{S}}}}
\newcommand{\Wk}[0] { \vekt{W}_k}
\newcommand{\Vk}[0] { \vekt{V}_k}
\newcommand{\vektil} [1] { \vekt{\tilde{#1}}}
\newcommand{\foralltime}[0]{\forall t \ge 0}
\newcommand{\dniqn}[1]{DN-IQN$_{#1}$}
\newcommand{\marktext}[1]{\textit{#1}}

\newcommand{\Vhrulefill}[0] {\leavevmode\leaders\hrule height 0.7ex depth \dimexpr0.4pt-0.7ex\hfill\kern0pt}

\newcounter{para}
\newcommand \mypara{\refstepcounter{para} \par \noindent\textit{Remark \thepara:\space}}
\newcommand{\alphaRN}[0]{\alpha^{RN}}

\newcommand{\reldiff}[1]{\fontsize{8}{9.6}{\textcolor{gray}{$#1$}}}
\newcommand{\iter}[1]{$\mathbf{#1}~$}
\newcommand{\NewtonIter}[1]{\textcolor{darkgray}{$\mathbf{#1}~$}}
\newcommand{\diverged}[0]{\multirow{2}{*}{\LARGE{\textbf{--}}}}

\newcommand{\E}[1]{\mathrm{E}{#1}}

\begin{document}

\title{A Robin-Neumann Scheme with Quasi-Newton Acceleration for Partitioned Fluid-Structure Interaction}

\author[1]{Thomas Spenke*}

\author[1]{Michel Make}

\author[1]{Norbert Hosters}

\authormark{Spenke \textsc{et al}}

\address{\orgdiv{Chair for Computational Analysis of Technical Systems (CATS), Center for Simulation and Data Science (JARA-CSD)}, \orgname{RWTH Aachen University}, \orgaddress{\country{Germany}}}

\corres{*Thomas Spenke, Chair for Computational Analysis of Technical Systems (CATS), Center for Simulation and Data Science (JARA-CSD), RWTH Aachen University, Germany. \email{spenke@cats.rwth-aachen.de}}

\abstract[Summary]{
The Dirichlet-Neumann scheme is the most common partitioned algorithm for fluid-structure interaction (FSI)
and offers high flexibility concerning the solvers employed for the two subproblems.
Nevertheless, it is not without shortcomings:
To begin with,
the inherent added-mass effect 
often destabilizes the numerical solution severely.
Moreover,
the Dirichlet-Neumann scheme cannot be applied to
FSI problems in which an incompressible fluid is fully enclosed by Dirichlet boundaries, as it is incapable of satisfying the volume constraint. 

In the last decade, interface quasi-Newton methods have proven to control the added-mass effect and 
substantially speed up convergence
by adding a Newton-like update step to the Dirichlet-Neumann coupling.
They are, however, without effect on the incompressibility dilemma.
As an alternative, the Robin-Neumann scheme
generalizes the fluid's boundary condition
to a Robin condition 
by including the Cauchy stresses.
While this modification in fact successfully tackles both drawbacks of the Dirichlet-Neumann approach,
the price to be paid is a strong dependency on the Robin weighting parameter, 
with very limited a priori knowledge about good choices.

This work proposes a strategy to
merge these two ideas and benefit from their combined strengths.
The resulting 
quasi-Newton-accelerated Robin-Neumann scheme
is
compared to both Robin- and Dirichlet-Neumann variants.
The numerical tests demonstrate that
it does not only provide faster convergence,
but also massively reduces the influence of the Robin parameter,
mitigating the main drawback of the Robin-Neumann algorithm.}


\jnlcitation{\cname{%
\author{T. Spenke},
\author{M. Make}, and
\author{N. Hosters}}
(\cyear{2022}),
\ctitle{A Robin-Neumann Scheme with Quasi-Newton Acceleration for Partitioned Fluid-Structure Interaction}, \cjournal{Int J Numer Methods Eng.}, \cvol{2022}.}
\maketitle


\section{Introduction}

Partitioned solution schemes for multi-physics problems
are widely used
in modern computational mechanics due to their modularity:
Treated as black boxes, the distinct solvers employed for the subproblems are coupled only via the exchange of interface data.
In fluid-structure interaction,
the prevalent partitioned algorithm is the \textit{Dirichlet-Neumann (DN)} scheme,
which passes the fluid loads at the interface
as a Neumann condition
to the structural solver, before the resulting deformation state
imposes a Dirichlet condition on
the flow problem.
Despite its popularity, the Dirichlet-Neumann scheme comes with two significant drawbacks addressed in this work:
(1) 
Its pronounced sensitivity to the \textit{added-mass effect} 
not uncommonly causes the coupling iteration to diverge,
preventing
any numerical solution
\cite{causin2005added,forster2007robust,forster2007artificial,van2009added}.
(2) It fails to work for FSI simulations with a fully-enclosed incompressible fluid, i.e.,
if the velocity field is prescribed by Dirichlet conditions on all boundaries,
such as for example the inflation of a water balloon.
In that case, the fluid volume is uniquely defined by the structural deformation,
which does not account for the flow's incompressibility,
causing what is referred to as the \textit{incompressibility dilemma} \cite{kuttler2006solution,Bogaers2015}.\\
%

Tackling the added-mass instability, \textit{interface quasi-Newton (IQN)} methods
have been a vibrant field of research in recent years and grown to be a key part in modern coupling libraries \cite{Bungartz2016},
%
%
as they both stabilize and accelerate partitioned schemes \cite{degroote2010performance,Bogaers2014,lindner2015comparison}.
Identifying the converged time step solution as a fixed point of the coupling loop,
their key idea is to add a Newton-like update step of the exchanged data fields, e.g., the interface deformation.
%
%
As the inverse Jacobian required for Newton's method is typically inaccessible, it is approximated instead.

After pioneering works in the field \cite{gerbeau2003quasi,van2005interface}, 
the 
interface quasi-Newton 
\textit{inverse least-squares (ILS)} method by Degroote et al. \cite{Degroote2009} marked an important milestone,
as it introduced the least-squares approximation of the inverse Jacobian
that has become standard today.
Initially, it relied only on input-output data pairs collected within 
%
the current time step.
Follow-up research 
revealed, however, that
using past time step data is very advantageous, but also challenging:
An explicit reutilization of data pairs may suffer from rank deficiency and the dependency on the number of reused time steps
- although both effects can be alleviated by proper filtering techniques \cite{haelterman2016improving,davis2022enhancing}.
An implicit incorporation via an updated Jacobian avoids these issues, but
typically brings along a costly explicit Jacobian \cite{Bogaers2014,lindner2015comparison}.
However, recent formulations circumventing this drawback were presented by
Scheufele and Mehl \cite{scheufele2017robust} as well as Spenke et al. \cite{spenke2020multi}.
Aside from reusing past data,
an interesting and rather new strategy is
to enhance the Jacobian approximation with a surrogate model,
obtained for example from simplified physics or coarser discretizations \cite{delaisse2022surrogate, demeester2021efficient}.
%
%
Furthermore,
R\"uth et al. \cite{ruth2021quasi} proposed a combination with waveform iterations to achieve higher-order convergence in time.\\

%
While update steps of the interface data, like quasi-Newton methods, 
can overcome the added-mass instability,
they are
without effect on the incompressibility dilemma,
as it is inherent to the Dirichlet-Neumann communication pattern 
itself.
One countermeasure is therefore to change this very pattern.
In 2008, Badia et al. \cite{badia2008fluid} 
as well as
Nobile and Vergara \cite{nobile2008effective} first proposed the usage of Robin-based schemes:
Rather than transferring either Dirichlet or Neumann interface data, 
these approaches make use of
Robin conditions to linearly combine the two contributions.
%
Bearing in mind the special cases of either contribution being zero,
the Dirichlet-Neumann scheme can in fact be sorted into
the more general family of Robin-Robin schemes.

In this work, however, another member is of particular interest, the \textit{Robin-Neumann (RN)} scheme:
While the fluid tractions are passed to the structural solver just as in the Dirichlet-Neumann case,
both the structure's deformation and tractions are returned to the fluid problem,
forming 
a Robin boundary condition.
It can be pictured as adding some numerical permeability to the FSI interface, 
allowing for artificial fluid fluxes that vanish when convergence is reached. 
The capability of temporarily violating kinematic continuity not only counteracts the added-mass effect,
but also frees the Robin-Neumann scheme from the incompressibility dilemma, as fully-enclosed cases
no longer have Dirichlet boundaries only.
Unfortunately, however, the Robin-Neumann scheme also introduces new drawbacks. In particular,
its performance is strongly governed by the Robin parameter that controls the weighting of Dirichlet and Neumann contributions;
and although tuning this parameter has been studied for various simplified FSI problems \cite{gerardo2010analysis,fernandez2013explicit, Cao2018,cao2021spatially},
efficient choices are in general
problem-dependent and difficult to find a priori.\\


Both interface quasi-Newton methods and the Robin-Neumann scheme
are preferable to the plain Dirichlet-Neumann approach in slightly different aspects;
but to the best of the authors' knowledge, the two approaches have never been combined.
This work proposes a novel Robin-Neumann scheme
with quasi-Newton acceleration
that merges the strengths of both approaches.
In particular, it 
allows to benefit from 
IQN methods
for FSI simulations with fully-enclosed incompressible fluids, 
for which the Dirichlet-Neumann scheme 
is inapplicable.
%
Interestingly, the proposed strategy not only improves the Robin-Neumann scheme's stability and convergence,
but also substantially reduces the dependency on the Robin parameter.
%
Beyond that, 
its potential performance gain over the Dirichlet-Neumann scheme with IQN update for non-enclosed cases
is demonstrated.

Aside from the quasi-Newton update, 
a direct feedback of fluid loads is introduced,
which renders any explicit computation of structural Cauchy stresses unnecessary
and increases the Robin-Neumann scheme's usability for black-box solvers.\\

This work is outlined as follows:
Section \ref{Sec:ProblemStatement} presents the governing equations,
before Section \ref{Sec:PartitionedFSI} focuses on partitioned algorithms for fluid-structure interaction,
including 
interface quasi-Newton methods and the Robin-Neumann coupling.
Based on that, the new \textit{Robin-Neumann quasi-Newton (RN-QN)} scheme is
proposed in Section \ref{Sec:RN-IQN}.
Its performance is demonstrated in Section \ref{Sec:Results}
via numerical test cases, 
covering both fully-enclosed and open fluid-structure interaction problems.


\section{Problem Statement} \label{Sec:ProblemStatement}

In fluid-structure interaction, 
a fluid domain $\Omega_t^f \subset \mathbb{R}^{\nsd}$ 
and a structural body $\Omega_t^s \subset \mathbb{R}^{\nsd}$
share a boundary 
$\Gamma_t^{fs} = \partial \Omega_t^f \cap \partial \Omega_t^s$.
The subscript $t$ indicates the time level, while $\nsd$ is the number of spatial dimensions.
%
This section briefly presents the two subproblems,
their numerical solution,
and the coupling conditions arising at the interface $\Gamma_t^{fs}$.

\subsection{Fluid Problem} \label{SubSec:FluidProblem}

The flow problem is governed by the unsteady \textit{Navier-Stokes equations} for an incompressible fluid,
\begin{subequations}
	\begin{alignat}{2}
		\rho_f \left( \frac{\partial \vel}{\partial t} + \vel \cdot \boldsymbol{\nabla} \vel - \vekt{f}_f \right) - \boldsymbol{\nabla}  \cdot \CauchyStress_f &= \vekt{0}	\qquad	&& \text{in} ~\Omega_t^f ~~\foralltime \,,\\
		\boldsymbol{\nabla}  \cdot \vel \,&= 0 && \text{in} ~\Omega_t^f ~~\foralltime \,, 
	\end{alignat}
\end{subequations}
that are solved for the fluid velocity $\vekt{u}_f(\vekt{x},t)$ and the pressure $p_f(\vekt{x},t)$. Therein,
$\rho_f$ is the constant fluid density, while $\vekt{f}_f$ denotes the resultant of all body forces per unit mass of fluid.
Assuming a Newtonian fluid 
with dynamic viscosity $\mu_f$,
the Cauchy stress tensor $\CauchyStress_f$
is given by:
%
$\CauchyStress_f( \vel, p_f) = - p_f \vekt{I} + \mu_f \left(  \nabla \vel + (\nabla \vel)^T \right)$,
with the unit matrix $\vekt{I}$.
%

The problem is closed by setting an initial flow field $\vel (\vekt{x},t=0)$
and the following boundary conditions
on the Dirichlet boundary $\Gamma_{D,\,t}^{f}$, the Neumann part $\Gamma_{N,\,t}^{f}$,
and a Robin section $\Gamma_{R,\,t}^{f}$, 
with $\Gamma_{D,\,t}^{f} \cap \Gamma_{N,\,t}^{f} \cap \Gamma_{R,\,t}^{f} = \partial \Omega^f_t$:
\begin{subequations}
	\begin{alignat}{2}
		\vel =& ~\vekt{g}_f  && \qquad \text{on} ~ \Gamma_{D,\,t}^{f} ~~\foralltime \,, \\
		\CauchyStress_f  \, \vekt{n}_f =& ~\vekt{h}_f  && \qquad \text{on} ~ \Gamma_{N,\,t}^{f} ~~\foralltime \,, \label{NeumannBcFluid} \\
		\alphaRN \vel+ \CauchyStress_f  \, \vekt{n}_f =& ~\vekt{h}_f + \alphaRN \vekt{g}_f  && \qquad \text{on} ~ \Gamma_{R,\,t}^{f} ~~\foralltime \,,   \label{RobinBcFluid}
	\end{alignat}
\end{subequations}
with the prescribed velocity $\vekt{g}_f$ and tractions $\vekt{h}_f$ \cite{hostersspline}.
Note that the Robin condition with its weighting factor $\alphaRN$ will contribute to both left- and right-hand side of the discrete system
due to the unknown $\vel$.\\

The fluid problem is solved by the in-house solver XNS, using P1P1 finite elements 
with \textit{Galerkin/Least-Squares (GLS) stabilization}
in space \cite{Pauli2017,donea2003finite} and a BDF1 scheme in time \cite{forti2015semi}.
The ALE mesh is adapted to the deforming domain via the \textit{linear elastic mesh-update method (EMUM)} \cite{johnson1994mesh,behr2002free};
its velocity is determined using a first-order finite difference scheme, in line with the BDF1 integration of the flow problem \cite{forster2006geometric}.

\subsection{Structural Problem}

The structural displacement field $\vekt{d}_s(\vekt{x},t)$ is computed from the dynamic balance of inner and outer stresses.
Using a total Lagrangian viewpoint,
it is formulated in the undeformed configuration, indicated for all affected quantities by the subscript $0$,
such as $\Omega^s_0$ or $\boldsymbol{\nabla}_0$.
%
The resulting equation of motion reads 
\begin{alignat}{2}
	\rho_s \frac{d^2 \vekt{d}_s}{dt^2} &= \boldsymbol{\nabla}_0 \cdot \left( \vekt{S} \vekt{F}^T \right) + \vekt{b}_s \qquad &&\text{in } \Omega_0^s ~~\foralltime\,,
\end{alignat}
where $\rho_s$ denotes the material density and $\vekt{b}_s$ the resultant of all body forces per unit volume. 
The deformation gradient $\vekt{F}$ relates the
2nd Piola-Kirchhoff stresses $\vekt{S}$ to the Cauchy stress tensor $\CauchyStress_s$, with $\vekt{S} = \det (\vekt{F}) ~ \vekt{F}^{-1}  \, \CauchyStress_s  \, \vekt{F}^{-T}$.
As constitutive equation,
the St. Venant-Kirchoff material model provides 
the linear stress-strain law
$\vekt{S}= \lambda_s \text{tr} \left( \vekt{E} \right) + 2 \mu_s \vekt{E}$,
with the Lam\'e constants $\lambda_s$, $\mu_s$ 
and the Green-Lagrange strains $\vekt{E} := \frac{1}{2} \left( \vekt{F}^T \vekt{F} - \vekt{I} \right)$ \cite{bathe2006finite}.

Aside from an initial deformation $\vekt{d}_s(\vekt{x},t=0)$,
Dirichlet and Neumann conditions are defined to close the problem. They set
the displacements $\vekt{g}_s$ on $\Gamma_{D,0}^s  \subset \partial \Omega^s_0$ 
and the tractions $\vekt{h}_s$ on $\Gamma_{N,0}^s \subset \partial \Omega^s_0$,  with the outer unit normal  $\vekt{n}_{s,0}$:
\begin{subequations}
	\begin{alignat}{2}
		\vekt{d}_s =& ~ \vekt{g}_s 		 && \qquad \text{on }  \Gamma_{D,\,0}^{s} ~~\foralltime \,, \\
		\vekt{F \,S} \, \vekt{n}_{s,0} =& ~ \vekt{h}_s		 && \qquad \text{on }  \Gamma_{N,\,0}^{s} ~~\foralltime \,.
	\end{alignat}
\end{subequations}

The structural subproblem is numerically solved by the in-house finite-element code FEAFA using 
\textit{isogeometric analysis (IGA)} \cite{hughes2005isogeometric,cottrell2009isogeometric} in space and a \textit{generalized-$\alpha$} scheme in time \cite{chung1993time,erlicher2002analysis}.

\subsection{Coupling Conditions} \label{Sec:CouplingConditions}

It is the very essence of multi-physics simulations that the subproblems cannot be solved independently.
For fluid-structure interaction, the solution fields are connected at the FSI interface $\Gamma_t^{fs}$  via the following coupling conditions:
\begin{enumerate}
	\item[$\bullet$] The \textit{kinematic} condition states the continuity of displacements,
	\begin{alignat}{2}
		\vekt{d}_f  &= \vekt{d}_s	\qquad &&\text{on } \Gamma_t^{fs} ~~\foralltime\,,
	\end{alignat}
	which analogously implies the equality of velocities $\vekt{u}_f=\vekt{u}_s$ and accelerations $\vekt{a}_f=\vekt{a}_s$, too.
	\item[$\bullet$] In keeping with Newton's third law,
	the \textit{dynamic} coupling condition requires the equality of interface tractions:
	\begin{alignat}{2}
		\CauchyStress_f  ~ \vekt{n}_f  &= \CauchyStress_s  ~ \vekt{n}_s  	\qquad &&\text{on } \Gamma_t^{fs} ~~\foralltime \,.
	\end{alignat}	
	Therein, 
	$\vekt{n}_f$ and $\vekt{n}_s$ denote the associated unit normal vectors (with $\vekt{n}_f = - \vekt{n}_s$).
\end{enumerate}

Satisfying these coupling conditions for every time $t$ in a continuous manner ensures the conservation of mass, momentum, and mechanical energy
over the FSI boundary \cite{kuttler2006solution,makespline}.


\section{Partitioned Fluid-Structure Interaction} \label{Sec:PartitionedFSI}

Following a partitioned FSI coupling approach, 
the two subproblems, i.e., fluid and structure, are addressed by two distinct solvers,
that are connected only via the exchange of interface data.
%
While this strategy features a high flexibility and modularity
regarding the two solvers,
their communication requires some additional considerations:
\begin{enumerate}
	\item[$\bullet$]
	In general, the spatial discretizations do not
	match at the FSI boundary, so that
	the transfer of interface data requires a
	conservative projection.
	In this work, we employ a spline-based 
	variant of the \textit{finite interpolation elements (FIE)} method, see
	Hosters et al. \cite{hostersspline,hosters2018fluid} and Make \cite{makespline}.
	For the sake of simplicity, however, this \textit{spatial coupling} will be neglected in the following 
	as it does not interfere with the methods discussed.
	\item[$\bullet$]
	The interdependency between the two subproblems in general requires an iterative procedure 
	to find a consistent solution of the coupled problem.
	This is referred to as \textit{strong (temporal) coupling} \cite{hostersspline,makespline,degroote2011multi}.
\end{enumerate}

\subsection{Dirichlet-Neumann Scheme} \label{Sec:DN}

\begin{figure}[t]
	\centerline{\includegraphics[width=0.68\textwidth]{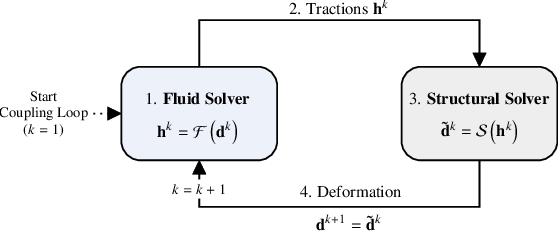}}
	\caption{The coupling loop of the Dirichlet-Neumann scheme, iterated until convergence for every time step.\label{Fig:DN}}
\end{figure}

The most widespread strongly-coupled algorithm for FSI problems is the \marktext{Dirichlet-Neumann (DN)} scheme.
Its name stems from using the dynamic coupling condition to pass the fluid tractions as a \marktext{Neumann} boundary condition to the structure,
before returning the resulting interface deformation -- in accordance with kinematic continuity -- as a \marktext{Dirichlet} condition to the fluid field\footnote{
	The Dirichlet condition of the fluid problem requires a velocity rather than a displacement field.
	In this work, the mesh is first adjusted to the interface deformation, before the fluid motion is set to the resulting
	mesh velocity.}.
The procedure is illustrated in Figure \ref{Fig:DN}.

To simplify notation,
the subscripts $s$ and $f$ of the interface deformation $\vekt{d}$ 
and the fluid tractions $\vekt{h}$, respectively, are dropped henceforth.
In every coupling iteration $k$, the following steps are repeated until convergence is reached:

\begin{enumerate}
	\item The fluid solver determines the tractions 
	$\vekt{h}^k = \fluidSolver(\vekt{d}^k)$ 
	based on the current deformation $\vekt{d}^k$.
	\item The tractions $\vekt{h}^k$ are passed to the structural solver, which
	\item computes the resulting interface deformation $\vekt{\tilde{d}}^k=\structSolver(\vekt{h}^{k})$.  
	\item The deformation $\vektil{d}^k$ is sent  to the flow solver and 
	will be its input 
	in the next iteration\footnote{ 
		The distinction between the output $\vektil{d}^k$ 
		of the structural solver and the deformation field
		$\vekt{d}^{k+1}$ used as boundary condition in the flow problem is a
		preparation 
		for the update step introduced in Section \ref{Subsec:UpdateTechniques}.}
i.e., $\vekt{d}^{k+1}=\vektil{d}^k$.
\end{enumerate}
The time step is considered converged if the following two criteria are fulfilled:
\begin{enumerate}[(I)]
\item Interpreting the coupling loop as a fixed-point iteration \cite{degroote2010performance,kuttler2008fixed},
both subproblem solutions 
(and the interface data)
have to stay virtually unchanged within one coupling iteration, i.e.,
their  relative changes have to be lower than some bound $\varepsilon_{Coupling}$. 
\item 
%
For accurate results, the 
residuals of the two nonlinear subproblems have to satisfy their own
convergence criterion $\varepsilon_{Problem}$ before going on to the next time step.
\end{enumerate}

In the simulations of Section \ref{Sec:Results}, 
both solvers are iterated to convergence in every call
to make sure the comparison of coupling iterations per time step is
not compromised by the solvers' internal Newton loops.
%
%
This strategy inherently fulfills condition (II),
but typically causes a significant overhead in computation time,
which is why
Spenke et al. \cite{spenke2021performance} propose to run just one Newton iteration per solver call,
unless convergence criterion (I) is already met.

\subsection{Added-Mass Instability and Incompressibility Dilemma} \label{SubSubSec:AddedMassDilemma}

Being a partitioned algorithm, the Dirichlet-Neumann scheme suffers from the {artificial \marktext{added-mass effect}:
Due to calling the solvers in a staggered manner,
in some coupling iteration(s) the subproblem solutions are inevitably computed with boundary conditions at 
$\Gamma^{fs}_t$
that differ from the converged state still to be found.
For instance, the flow field is generally updated based on an inexact interface deformation.
In case this deformation is overestimated, 
kinematic continuity causes an exaggerated fluid acceleration at the interface, 
resulting in
excessive inertia terms 
that act as an additional artificial fluid mass on the structure.
This added-mass instability is in particular increasing with the density ratio between the fluid and the structure $\rho_f / \rho_s$.
For more information on the added-mass effect, we recommend the works by Causin et al. \cite{causin2005added}, F\"orster et al. \cite{forster2007robust,forster2007artificial}, and van Brummelen \cite{van2009added}.

Aside from the added-mass instability, 
the Dirichlet-Neumann scheme has a second weakness,
%
%
referred to as 
\textit{incompressibility dilemma}:
It does not work for FSI problems in which an incompressible fluid is fully enclosed by Dirichlet conditions for the velocity, such as walls or prescribed inflows.
For such a problem, the fluid volume is uniquely defined by the interface deformation;
however, the structural solver is unaware of the fluid's incompressibility constraint,
leaving the coupling procedure bound to fail \cite{kuttler2006solution, Bogaers2015}.

\subsection{Update of Coupling Data} \label{Subsec:UpdateTechniques}

One countermeasure against the added-mass instability 
is to
modify the coupling data in an update step
before passing it on to the other solver.
Typically, 
%
the assignment $\vekt{d}^{k+1} =  \vektil{d}^k$ in step 4 of the Dirichlet-Neumann scheme is replaced by some update step $\vekt{d}^{k+1} = \mathcal{U}(\vektil{d}^k)$.
%
%
But as the concept is applicable to other interface data too,
we will refer to
some generic interface field $\vektil{x}^k \in \mathbb{R}^m$
computed in iteration $k$,
that is updated to the next iteration's input by $\vekt{x}^{k+1} = \mathcal{U}(\xtil^k)$.
The associated fixed-point residual $\vekt{R}^k =\xtil^k - \vekt{x}^k$
quantifies the change of $\vekt{x}$ within coupling iteration $k$.

The simplest update is a \marktext{relaxation} of the coupling data,
\begin{align}
	\vekt{x}^{k+1} = \mathcal{U}_{Relax}(\xtil^k) =  \omega \, \xtil^k + (1-\omega) \, \vekt{x}^k \, , \label{Eqn:Relaxation}
\end{align}
where a relaxation factor $0<\omega < 1$ is required to increase stability 
(``under-relaxation''). 
%
Unfortunately, a fixed $\omega$ often has to be chosen very small to ensure stability for all time steps, bringing along a drastic decrease in efficiency.
Therefore,
\marktext{Aitken's dynamic relaxation} \cite{kuttler2008fixed, irons1969version} adapts the relaxation factor in every coupling iteration by
\begin{align}
	\omega_k = - \omega_{k-1} \frac{ (\vekt{R}^{k-1})^T (\vekt{R}^k - \vekt{R}^{k-1} ) }{ \| \vekt{R}^k - \vekt{R}^{k-1} \|_2^2 } 
\end{align}
before the update step (\ref{Eqn:Relaxation}).
Despite its simplicity,
%
Aitken's relaxation often significantly speeds up convergence.
It can be interpreted as a simplified version of 
\textit{interface quasi-Newton (IQN)} methods.

\subsection{Interface Quasi-Newton Methods} \label{SubSec:IQN}

Identifying the converged time step solution as a root of the fixed-point residual $\vekt{R}$, 
convergence could be accelerated
by using Newton's method as an update step, i.e.,
\begin{align}
	\vekt{x}^{k+1} =  \mathcal{U}_{Newton}(\xtil^k) = \vekt{\tilde{x}}^k - \vekt{J}^{-1}_R (\vekt{\tilde{x}}^k)~\vekt{R}^k \,.   \label{Eqn:NewtonUpdate}
\end{align}
As the inverse Jacobian $\vekt{J}^{-1}_R := (d \vekt{R}/d\vektil{x})^{-1}$ is typically not accessible
due to the required derivatives of the two solvers, 
interface quasi-Newton methods instead rely on
an approximation $\Jac \approx \vekt{J}^{-1}_{\tilde{R}} \in \mathbb{R}^{m \times m}$.

It is based on multi-dimensional finite differences,
forming the required
data pairs ($\Delta \vektil{x}$, $\Delta \vekt{R}$) 
from the intermediate fields $\vektil{x}^i$ and $\vekt{R}^i$ of the $k$ coupling iterations already performed for the current time step.
They are stored in the matrices $\vekt{V}_k \in \mathbb{R}^{m \times k}$ and $\vekt{W}_k \in \mathbb{R}^{m \times k}$ \cite{Degroote2009,scheufele2017robust}:
\begin{subequations}
	\begin{alignat}{7}
		\vekt{V}_k &= \big[ \Delta \vekt{R}_{k-1}^k, ~ &&\Delta \vekt{R}_{k-2}^{k-1},~ && \cdots, ~ &&\Delta \vekt{R}_0^1 && \big] 
		&&\text{with }
		\Delta \vekt{R}_i^j && = \vekt{R}^j - \vekt{R}^i   ~, \\
		\vekt{W}_k &= \big[ \Delta \vekt{\tilde{x}}_{k-1}^k ~,  &&\Delta \vekt{\tilde{x}}_{k-2}^{k-1} \,, && \cdots, && \Delta \vekt{\tilde{x}}_0^1   && \big] 
		\qquad 
		&& \text{with }
		\Delta \vekt{\tilde{x}}_i^j &&= \vekt{\tilde{x}}^j - \vekt{\tilde{x}}^i \,.
	\end{alignat}  \label{Eqn:InputOutputMatrices}
\end{subequations} 
The collected data poses the constrained optimization problem
\begin{align}
	\min	\|\, \Jac  \|_F \quad \text{subject to} \quad \Jac ~ \vekt{V}_k = \vekt{W}_k  \,. \label{Eqn:OptimizationILS}
\end{align}
While the linear system alone would be underdetermined ($m >> k$),
the Frobenius norm minimization provides the unique solution
$\Jac  = \vekt{W}_k \left( \vekt{V}_k^T \vekt{V}_k \right) ^{-1} \vekt{V}_k^T$ \cite{lindner2015comparison,scheufele2017robust},
that
is inserted into Equation (\ref{Eqn:NewtonUpdate}) to obtain
the quasi-Newton update
\begin{align}
	\vekt{x}^{k+1} &= \mathcal{U}_{IQN}(\xtil^k) =
	\xtil^k +\Jac \left( - \vekt{R}^k\right) =
	\xtil^k + \vekt{W}_k  \underbrace{  \left( \vekt{V}_k^T \vekt{V}_k \right) ^{-1} \vekt{V}_k^T   ~\left( - \vekt{R}^k \right)  }_{:=  \boldsymbol{\alpha}^k}\label{Eqn:IQN_ILS_Update} 
	= \xtil^k + \vekt{W}_k \boldsymbol{\alpha}^k
	\,. 
\end{align}
Since we do not need the Jacobian explicitly,
this update is evaluated by 
solving the least-squares problem
\begin{align} 
	\min_{\boldsymbol{\alpha}^k \in \mathbb{R}^k} \| \vekt{V}_k \boldsymbol{\alpha}^k + \vekt{R}^k \|_2   \label{Eqn:LSalpha}
\end{align}
for the vector $\boldsymbol{\alpha}^k \in \mathbb{R}^{k}$,
e.g., using a QR decomposition via Householder reflections. 
In the first coupling iteration, when no data pairs are available yet,
%
a relaxation step is used.

The technique outlined so far is the interface quasi-Newton \textit{inverse least-squares (ILS)} method \cite{Degroote2009}.
While its basic form
%
only considers data collected during the current time step, the effectiveness of IQN methods is significantly improved by using data from past time steps too.
A straightfoward option is to explicitly include the data pairs of the $q$ most recent time steps in $\Vk$ and $\Wk$. 
But although proper filtering techniques can alleviate the issue,
good choices of the parameter $q$ are problem-dependent due to numerical challenges like bad conditioning, rank-deficiencies, and contradicting information \cite{haelterman2016improving}. 

As an alternative, multi-vector quasi-Newton methods reuse past data in an implicit manner \cite{Bogaers2014,scheufele2017robust}:
Formulating the new inverse Jacobian $\JacN{n+1}$ as an update from that of the previous time step $\JacN{n}$, 
the Jacobian in the Frobenius norm minimization of
Equation (\ref{Eqn:OptimizationILS}) 
is replaced by the increment:
$\min  \| \JacN{n+1} - \JacN{n} \|_F$.
%
%
This results in the multi-vector approximation
$\JacN{n+1} = \JacN{n}  + \left( \Wk - \JacN{n}  \, \Vk \right)\left( \Vk^T \Vk \right)^{-1} \Vk^T$
and the update step
\begin{align}
	\vekt{x}^{k+1}
	= \mathcal{U}_{IQN}(\xtil^k ) 
	=  \xtil^k - \JacN{n} \vekt{R}^k + \left( \Wk - \JacN{n} \, \Vk \right) \underbrace{  \left( \Vk^T \Vk \right)^{-1} \Vk^T (- \vekt{R}^k) }_{\boldsymbol{\alpha}^k}
	=
	\xtil^k - \JacN{n} \vekt{R}^k + \left( \vekt{W}_k - \JacN{n} \vekt{V}_k \right) \boldsymbol{\alpha}^k \,. \label{Eqn:IMVLS_implicitUpdate} 
\end{align}
Solving the same least squares problem as in Equation (\ref{Eqn:LSalpha}) for $\boldsymbol{\alpha}^k$,
the \marktext{implicit multi-vector least squares (IMVLS)} variant allows to evaluate this expression without any explicit Jacobian representation \cite{spenke2020multi}.

Aside from the references already given, 
the works
by Lindner et al. \cite{lindner2015comparison} or
Delaiss\'e et al. \cite{delaisse2021comparison}
are recommended for an overview of different IQN methods.

\subsection{Robin-Neumann Scheme} \label{SubSec:RN}

Despite the popularity of the Dirichlet-Neumann scheme,
using the dynamic coupling condition to pass data from the fluid to the structure and the kinematic one for the way back
is not without alternatives.
%
As the Dirichlet condition 
of the flow problem 
is critical
for both added-mass instability and the incompressibility dilemma,
the \marktext{Robin-Neumann (RN)} scheme replaces it by 
a Robin condition that forms a linear combination of kinematic and dynamic continuity.
%
Inserting the structure's velocity $\vekt{u}_s =\partial \vekt{d}_s / \partial t$ and interface tractions $\vekt{T}_s \vekt{n}_s$ on $\Gamma^{fs}_t$,
Equation (\ref{RobinBcFluid}) can be rearranged to
\begin{align}
	\vekt{T}_f \vekt{n}_f  = \vekt{T}_s \vekt{n}_s + \alpha^{RN}  ~ \left( \frac{\partial \vekt{d}_s}{\partial t} - \vekt{u}_f \right) 	\qquad \text{on}~\Gamma^{fs}_t ~~  \foralltime \,, \label{RobinBC}
\end{align}
where the scalar parameter $\alpha^{RN} > 0$ controls the weighting of the two contributions.  In the limits, this condition reduces to a Neumann expression for $\alpha^{RN}=0$ and to 
a (weakly-imposed) Dirichlet condition for $\alpha^{RN} \to \infty$.

Typically, the tractions $\vekt{T}_s \vekt{n}_s$ originate from the structural solution.
But as their computation often requires
reconstruction techniques or might even be impossible for a black-box solver,
we propose a different strategy:
Due to the very nature of the structural problem as a dynamic balance of stresses,
the structure's Cauchy tractions 
at the interface
are already known
before even calling the solver:
They will 
balance the external loads, i.e.,
the fluid tractions 
passed to the structure\footnote{Technically, this equality only holds if the structural subproblem has fully converged. For other cases, it is still a very good approximation.}.
Consequently,
instead of explicitly computing any structural  stress,
these fluid tractions can directly be fed back into the Robin condition of the next iteration.
%
The resulting coupling strategy is illustrated in Figure \ref{Fig:RN}.

\begin{figure}[h]
	\centerline{\includegraphics[width=0.68\textwidth]{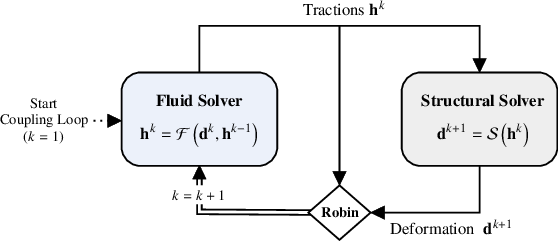}}
	\caption{Robin-Neumann scheme with a direct feedback of fluid tractions into the Robin condition.\label{Fig:RN}}
\end{figure}

In coupling iteration $k$, the Robin condition then reads
\begin{align}
	\vekt{T}_f \vekt{n}_f  =
	\vekt{h}^{k-1} + \alpha^{RN}  ~ \left( \frac{\partial \vekt{d}^k}{\partial t} - \vekt{u}_f \right) 
	\qquad \text{on}~\Gamma^{fs}_t ~~\foralltime \,. \label{RobinBCwithFeedback}
\end{align}

The Robin-Neumann scheme not only reduces the added-mass instability, 
but also resolves the incompressibility dilemma.
%
The key lies in weakening the Dirichlet condition of the flow problem,
in principle allowing for a violation of kinematic continuity, but penalizing it with the factor $\alphaRN$.
Accordingly, 
the velocity difference of fluid and structure
tends to zero
as the coupling converges.
To picture this, 
imagine the FSI interface is still moved with the structural deformation, but is no longer completely impermeable. 
Instead, an artificial flux through the boundary is permitted, so that the flow field always satisfies the incompressibility constraint.
This way,
fully-enclosed fluid problems are unproblematic,
since numerically, they are no longer of pure-Dirichlet type.
Concerning the added-mass effect, the artificial permeability 
caps the direct link between overestimated deformations and excessive inertia terms,
as the flow no longer ``blindly'' follows the interface;
this lowers the risk of
a reciprocal amplification
and increases stability.

Unfortunately, these strengths are put into perspective
by the main drawback of the Robin-Neumann scheme, i.e.,
its dependency on the Robin parameter.
%
While decreasing $\alphaRN$ stabilizes the coupling by weighting the Neumann contribution, it slows down convergence and 
%
reduces the solution quality,
in the sense that for a given convergence bound the remaining artificial flux increases.
For growing values of $\alphaRN$, in contrast, the accentuated Dirichlet part 
magnifies the added-mass instability. Moreover, it gradually reintroduces the incompressibility dilemma for fully-enclosed problems
by a stricter penalization of deviations from kinematic continuity.

Good choices of $\alpha^{RN}$ balance out these two opposing trends, but are very difficult to determine a priori.
Many very recommendable works
study the effect of the Robin parameter in detail,
like Badia et al. \cite{badia2008fluid}, 
Gerardo-Giorda et al. \cite{gerardo2010analysis}, 
Fern\'andez et al. \cite{fernandez2013explicit}, or
Cao et al. \cite{Cao2018,cao2021spatially}.
Analyzing simplified FSI problems,
for example potential or inviscid flows interacting with linear beam or membrane models,
%
they derive suggestions for choosing $\alpha^{RN}$,
including both constant values and spatially varying expressions \cite{cao2021spatially}.
Despite their undisputed importance,
these formulations
%
depend on a broad set of parameters, such as material properties, geometry, and time step size,
and typically have a limited applicability to more general problems.
Another option could be to 
choose the Robin parameter
by computing the structure's response to some test pressure loads a priori,
adapting the strategy proposed by Degroote et al. \cite{Degroote1000} for the artificial compressibilty coefficient.
%

All in all,
finding the best Robin parameter is still an ongoing research topic.
However, 
as the numerical results in Section \ref{Sec:Results} demonstrate,
the quasi-Newton acceleration proposed in this work
%
substantially reduces the impact of the Robin parameter,
and with it the importance of finding a good choice.
%
%
Similarly, a reduced parameter sensitivity was also observed by
Badia et al. \cite{badia2009robinkrylov},
when reformulating a Robin-Robin scheme as a preconditioned Richardson iteration over a Schur complement equation
to then exchange it for a Krylov iteration.\\

\mypara Although referring to tractions so far, 
passing stress tensors or forces from fluid to structure works analogously.
In our finite-element framework, 
we transfer consistent nodal forces,
which are
computed from integrating the tractions over the FSI interface \cite{hosters2018fluid}.
The same force term can be identified in the 
finite-element boundary integral of the Robin condition,
reading
(with the test functions $\vekt{w}_f$)
\begin{align}
	\int \limits_{\Gamma_t^{fs}}  \vekt{w}_f   \CauchyStress_f  \vekt{n}_f \, d \Gamma = 
	%
	\underbrace{\int \limits_{\Gamma_t^{fs}} \vekt{w}_f   \vekt{h}^{k-1} d \Gamma}_{\vekt{F}^{k-1}} + \int \limits_{\Gamma_t^{fs}} \vekt{w}_f  \alpha^{RN}   \left( \frac{\partial \vekt{d}^k}{\partial t} - \vekt{u}_f \right) d \Gamma  =
	\vekt{{F}}^{k-1}  + \int \limits_{\Gamma_t^{fs}}   \vekt{w}_f  \alpha^{RN} \left(  \frac{\partial \vekt{d}^k}{\partial t} - \vekt{u}_f \right) d \Gamma \, ,
\end{align}
so that the direct feedback of fluid loads is applicable to the nodal forces $\vekt{F}^k$, too. \label{Remark:ForceIntegral}


\section{Robin-Neumann Quasi-Newton (RN-QN) Scheme} \label{Sec:RN-IQN}

Despite being inherently different concepts, both the Robin-Neumann scheme and interface quasi-Newton methods can improve the coupling's stability and efficiency.
The main motivation of this work is to combine the two approaches and benefit from their respective strengths.
We therefore propose the following \marktext{Robin-Neumann quasi-Newton (RN-QN)} scheme illustrated in Figure
\ref{Fig:RN-IQN}:

\begin{figure}[h]
	\centerline{\includegraphics[width=0.68\textwidth]{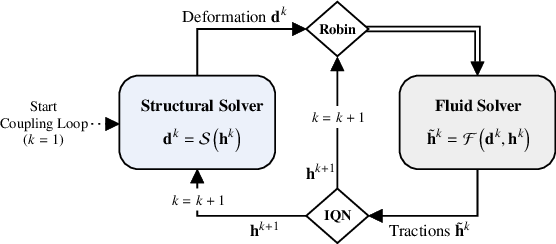}}
	\caption{Sketch of the proposed RN-QN scheme which combines the Robin-Neumann approach with an interface quasi-Newton acceleration. Key features are the quasi-Newton update of the fluid loads and their direct feedback into the fluid's Robin condition.\label{Fig:RN-IQN}}
\end{figure}

Essentially, the Robin-Neumann scheme 
as presented in Section \ref{SubSec:RN} 
is extended by an IQN method.
Due to the direct feedback of fluid tractions, however,
updating the interface deformation would cause
an inconsistency  between the Dirichlet and Neumann contributions of the Robin condition:
After the IQN update, the deformation would no longer correspond to the tractions the structural problem was solved for,
but to some stress state we have no chance of knowing without explicitly computing it.
To circumvent this issue, the quasi-Newton step of the RN-QN scheme modifies the fluid loads instead.
The resulting updated tractions $\vekt{h}^{k+1} = \mathcal{U}_{IQN}(\vektil{h}^k)$
enter both the structural solver and the fluid's Robin boundary condition in the next iteration.
Lastly, the order of the two solver calls has been switched,
%
as
the collection of input-output data slightly profits
from performing the quasi-Newton update in the end of a coupling iteration.
The new procedure is outlined in 
Algorithm \ref{Alg:RN-QN}:
Coupling iteration $k$ starts with the structural solver computing the deformation $\vekt{{d}}^k = \structSolver(\vekt{h}^k)$, 
%
which is
combined with the tractions $\vekt{{h}}^k$ to form a Robin condition.
With that, the flow solver determines new tractions $\vekt{\tilde{h}}^k = \fluidSolver( \vekt{d}^{k}, \vekt{h}^k)$.
After the quasi-Newton step $\vekt{h}^{k+1} = \mathcal{U}_{IQN}(\vekt{\tilde{h}}^k)$, 
the updated tractions $\vekt{h}^{k+1}$
are passed back to the structure and stored for
the next iteration's Robin term.
\begin{algorithm}[h!]
\caption{Pseudo-code of the new RN-QN scheme. The outlined procedure is repeated for every time step. }\label{Alg:RN-QN}
\begin{algorithmic}
	\For {\textbf{each coupling iteration} $k=1,\cdots$} \textbf{until converged}
		\State Call structural solver:   				\tabto{10cm} $\vekt{{d}}^k ~~~~ \,= \structSolver(\vekt{h}^k)$
	    \State {Pass deformation unaltered:}    \tabto{10cm} $\vekt{d}^{k}$ 
		\State{Form Robin condition:}  				 \tabto{10cm} $\CauchyStress_f \vekt{n}_f =  \vekt{h}^k + \alpha^{RN}  ~ \left( \frac{\partial \vekt{d}^{k} } {\partial t} - \vekt{u}_f \right)$
		\State{Call fluid solver:}   						\tabto{10cm} $\vekt{\tilde{h}}^k \quad ~= \fluidSolver( \vekt{d}^{k}, \vekt{h}^k)$
		\State{IQN update of fluid loads:} 			\tabto{10cm} $\vekt{h}^{k+1} ~~= \mathcal{U}_{IQN}(\vekt{\tilde{h}}^k)$
		\State{Pass resulting loads to structure and store for Robin condition:}  \tabto{10cm} $\vekt{h}^{k+1}$
	\EndFor
	\State Set initial tractions of next time step: \tabto{10cm}  $\vekt{h}^1 = \vekt{h}^{k+1}$ 
\end{algorithmic}
\end{algorithm}

\mypara Depending on which data field is sent,
the quasi-Newton step can equivalently be performed on tractions, stress tensors, or forces.
Note, however, that updating stress tensors
might affect the conditioning
of the least-squares problem
negatively
due to the typically very different scales of shear stresses and pressure.
For the numerical examples in Section \ref{Sec:Results}, the consistent nodal forces (see Remark \ref{Remark:ForceIntegral}) are fed into the IQN method. \\

\mypara \label{Remark:DN_IQN_F}
The idea to switch the order of solver calls and apply the quasi-Newton step to the
fluid loads rather than the interface deformation,
as done for the RN-QN scheme,
is just as well applicable to the Dirichlet-Neumann approach.
The performance analysis 
of different coupling schemes 
in Section \ref{Sec:Results} will therefore include this variant, labeling it \textit{\dniqn{f}}, as opposed to \textit{\dniqn{s}} for the classical update of deformations. \\


\mypara In this work, we characterize the Robin-Neumann scheme with a direct feedback of fluid loads as
suitable for black-box solvers, referring to the fact that no additional interface data fields 
(compared to the Dirichlet-Neumann approach) are exchanged
and no field equations have to be modified. The only further requirement is the flow solver's capability to handle a Robin condition, i.e., a nonlinear Neumann term.\\
%
%
%
Although various valuable works \cite{badia2008fluid,gerardo2010analysis,Cao2018}  link the fluid's Robin parameter to the characteristics of the structural subproblem,
this relation does not refer to any explicit data transfer,
%
but rather to a physical interpretation of $\alphaRN$.
%
%
Since the results in Section \ref{Sec:Results} show
the RN-QN scheme's
fast convergence
far off from physically motivated choices,
this work adapts the notion of $\alphaRN$ as a purely numerical quantity,
%
which does not compromise the black-box idea.\\

\mypara
To the best of the authors' knowledge,
this work is the first to propose a quasi-Newton-accelerated Robin-Neumann scheme.
However, Bogaers et al. \cite{Bogaers2015} applied a quasi-Newton update to the Dirichlet-Neumann approach 
with the interface artificial compressibility (DN-IAC) method.
Despite the similarities between RN and DN-IAC \cite{degroote2011similarity},
there are important differences to the RN-QN scheme:
While the Robin-Neumann scheme always satisfies incompressibility and only penalizes deviations from kinematic continuity,
the DN-IAC approach does the opposite. 
Moreover, it adds an additional source term to the continuity equation in the elements adjacent to the FSI interface and is thus arguably more invasive than adding a Robin boundary condition.
Lastly, this extra term does not consider viscous stresses, unlike the Robin traction part \cite{degroote2011similarity}.
%
Regarding the quasi-Newton update, Bogaers et al. rely on a block Newton technique
that approximates an own interface Jacobian for each solver,
rather than the inverse of the fixed-point operator's Jacobian.
While this approach yields comparable convergence rates,
%
the block quasi-Newton strategy requires the solution of two linear systems per coupling step and comes with a higher implementation effort \cite{Bogaers2014,uekermann2016partitioned},
in particular if the computational cost are supposed to grow linearly with the interface degrees of freedom to ensure the IQN method's efficiency for large-scale simulations \cite{spenke2020multi,delaisse2021comparison}.
Lastly, Bogaers et al. did not study the quasi-Newton update's effect on the parameter-dependency of the artificial compressibility method.


\section{Numerical Examples} \label{Sec:Results}

This section investigates the performance of the new quasi-Newton-accelerated 
Robin-Neumann scheme in two steps: 
First, its superiority over the plain Robin-Neumann scheme is shown for two enclosed-fluid problems, for which the Dirichlet-Neumann scheme is not applicable.
%
Afterwards, the RN-QN scheme is compared to the Dirichlet-Neumann scheme with IQN update for an open fluid configuration.\\

\mypara 
The computational cost per coupling iteration
is virtually the same for all partitioned schemes discussed in this work, i.e.,
Dirichlet-Neumann and Robin-Neumann variants with or without quasi-Newton update.
That is because the solver calls are expensive enough to
render the differences negligible, which is in line with literature \cite{scheufele2017robust,spenke2020multi}.
Consequently, in the following a scheme's efficiency will 
be assessed by the coupling iterations required.

\subsection{Inflating Balloon} \label{Sec:Ballon}

The first test case considers the inflating 2D water balloon depicted in Figure \ref{Fig:BalloonCase}: A flexible ring structure is filled with an incompressible fluid.
Due to a time-dependent in-/outflow condition at the interior fluid boundary,
the balloon will be in- and deflating \cite{hostersspline,makespline}. The parameters are listed in Table \ref{Tab:BalloonParameters}. 
Note in particular the high density ratio of $\rho_f / \rho_s = 1.0$, which hints at a pronounced added-mass instability.

\begin{figure}[h!]
	\centerline{\includegraphics[width=0.35\textwidth]{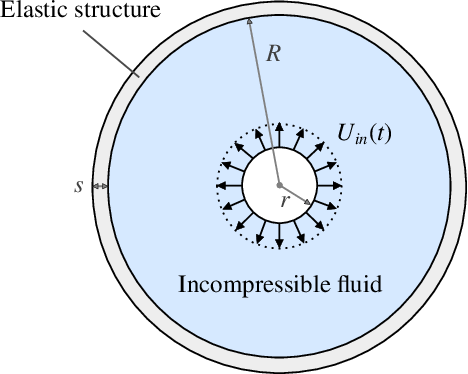}}
	\caption{Sketch of the inflating balloon case.}\label{Fig:BalloonCase}
\end{figure}

Since the symmetric inflow condition is known, the balloon volume, and hence the radius, can be computed analytically. Taking into account the geometrical error introduced by
approximating the circular FSI interface with $n$ linear segments, the analytical radius 
is:
\begin{align}
	R_{Polygon}(t) = \sqrt{R_0^2 + \frac{2 \pi r}{\frac{n}{2} \sin \frac{2 \pi}{n} }  \int_{0}^{t}  U_{in}(\bar{t}) \, d \bar{t} \,} ~.	\label{Eqn:PolygonRadius}
\end{align}

In this case, the interface is discretized by $n=60$ edges of the fluid mesh,
%
which in total consists of $2400$ triangular finite elements with $1260$ nodes.
%
%
The elastic structure is meshed by $40 \times 3 = 120$ nonlinear quadrilateral isogeometric elements, that are defined on a second-degree NURBS with $225$ control points.
The simulated time span of $t_{end}=2.0$ is divided into $n_{ts}=500$ steps of size $\Delta t=0.004$. 

As discussed in Section \ref{Sec:DN}, the convergence criterion takes both
the coupling loop and the Newton steps of the subproblems into account.
The bounds for this test case are $\varepsilon_{Coupling}=10^{-6}$
and 
$\varepsilon_{Problem}=10^{-10}$ for both problems.

\subsubsection{Results}

Due to the inflow over the boundary and the incompressibility constraint, the fluid domain is expanding, causing a widening of the elastic structure.
As soon as the boundary flux is switching to an outflow condition, the system starts shrinking again. Figure \ref{fig:BalloonSnapshots} visualizes this in- and deflation by
two snapshots and a plot of the balloon radius over time.

\begin{figure}[h!]
	\centerline{\includegraphics[width=1.0\textwidth]{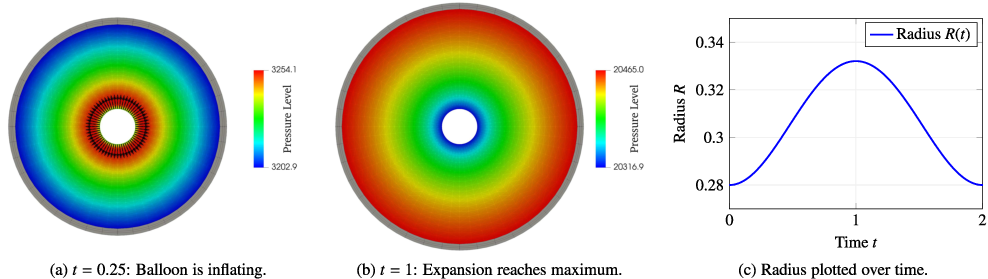}}
	\caption{Illustration of the balloon's in- and deflation.}\label{fig:BalloonSnapshots}
\end{figure}

The focus of this work, however, is less on the simulation results themselves
than on the performance of different FSI coupling strategies.
Due to the pure-Dirichlet character of the fluid problem, 
the Dirichlet-Neumann scheme is not applicable to this test case.
Instead, Table \ref{Tab:BalloonResults} compares 
multiple variants of the Robin-Neumann scheme: With or without quasi-Newton acceleration
and for various choices of the parameter $\alphaRN$.
For the RN-QN method, three different update techniques are listed:
(1) The IQN-IMVLS method with an implicit reutilization of past data, (2) the IQN-ILS approach without any past data, and (3) Aitken's dynamic relaxation.
Aside from the average number of coupling iterations required per time step,
the violation of the kinematic constraint is compared. It is quantified via
the relative artificial flux through the FSI interface
in the simulated time span $[0, t_{end}]$,
i.e.,
\begin{align}
	\varepsilon_{rel}
	= \int_{0}^{t_{end}} \frac{ \left| \, \dot{V}^{fs}(t) \, \right| }{V(t)} dt 
	%
	%
	\qquad \text{with} ~~ \dot{V}^{fs}(t) =  \int_{\Gamma_t^{fs}}    \left| \, \left( \frac{\partial \vekt{d}_s}{\partial t} - \vekt{u}_f \right) \cdot \vekt{n} \, \right| ~  d\Gamma  
	\quad \text{and} ~~ V(t) = V_0 + \int_0^t  \Sigma ~ \dot{V}( \bar{t})  \, d  \bar{t}     \,. \label{Eqn:SigmaRel}
\end{align}
This measure is chosen rather than the
averaged relative deviation from the analytical radius,
as the latter is dominated by the simulation's temporal discretization error and therefore
virtually the same for all runs
(about $4.0 \cdot 10^{-4}$).

The first row shows both the main strength and the biggest shortcoming 
of the Robin-Neumann scheme:
Although it is in principle capable of yielding efficient and accurate results, its performance 
massively depends on the choice of the Robin parameter.
Away from some acceptable range, small values lead to an impractically slow convergence;
picking $\alphaRN$ bigger than some critical bound, on the other hand, 
causes the coupling iteration to diverge altogether. 

In contrast,
the results of the new Robin-Neumann scheme with quasi-Newton acceleration
show only a limited dependency on $\alphaRN$,
in particular if past time step data is incorporated in the Jacobian approximation (IMLVS).
In fact, for the biggest part of the tested parameter range
%
it yields an almost constant convergence rate
that
even outperforms
the plain Robin-Neumann scheme with an optimal parameter.
Consequently, 
the further $\alphaRN$ moves away from the optimum,
the more substantial the relative speed-up of the new RN-QN coupling scheme gets.
%

Aside from performance, 
Table \ref{Tab:BalloonResults} 
also indicates how increasing the Robin parameter
improves the solution quality,
i.e., the conformity with kinematic continuity.
As expected for a penalty-enforced constraint,
the magnitude of the remaining artificial flux is solely determined by the choice of $\alphaRN$
and unaffected by the quasi-Newton update.

Comparing the update techniques employed within the RN-QN approach,
it should be noted:
(1) The convergence rate clearly benefits from reusing past time step data in the quasi-Newton update, e.g., via the IMVLS method, in particular for small Robin parameters.
(2) Despite being less effective than a quasi-Newton technique, Aitken's relaxation is still an easily implemented option to start with in case no IQN framework is available yet.

All in all, this test case clearly indicates that the new RN-QN scheme is preferable to the standard Robin-Neumann coupling,
as it strongly speeds up convergence and substantially reduces the dependency on the Robin parameter $\alphaRN$.

\subsection{Elastic Tube} \label{Subsec:Tube}

The next test case considers the cylindrical pipe 
filled with an incompressible fluid
depicted in Figure \ref{fig:PressureTubeCase}.
Due to a short excitation in the beginning,
a deformation wave will propagate through the elastic channel wall.
Note that unlike in similar cases from literature \cite{degroote2010performance,lindner2015comparison},
this excitation is not introduced via a pressure boundary condition.
Instead, a high inflow velocity is prescribed for a short time at one end of the pipe.
This modification allows to switch between two scenarios simply 
by changing the boundary condition at the opposite end of the pipe:
It will either (1) be closed to form a pure-Dirichlet problem
or (2) allow for free outflow via a natural Neumann condition.
\begin{figure}[h!]
	\centerline{\includegraphics[width=0.9\textwidth]{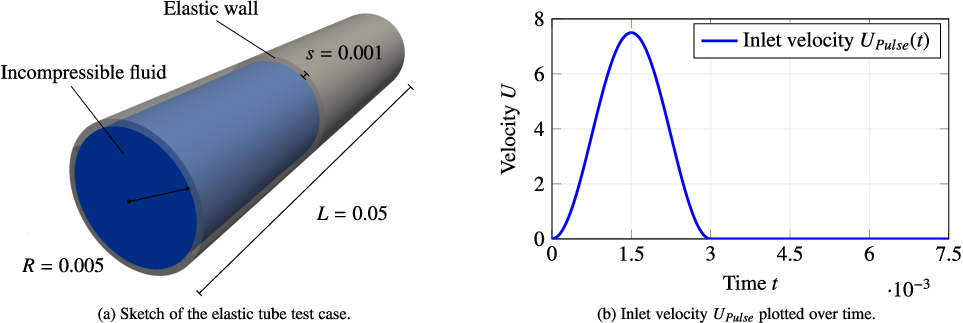}}	
	\caption{Elastic tube test case} \label{fig:PressureTubeCase}
\end{figure}

The tube has the inner radius $R=0.005$, wall thickness $s=0.001$, and length $L = 0.05$.
The fluid density is $\rho_f=1000.0$ and its dynamic viscosity $\mu_f=0.003$.
The elastic wall is characterized by the density $\rho_s=1000.0$, Young's modulus $E_s = 3.0 \cdot 10^5$,
and a Poisson ratio of $\nu_s=0.3$,
so that
the density ratio is again $\rho_f / \rho_s = 1.0$.

At the channel inlet, 
a parabolic inflow velocity profile with a time-dependent magnitude is prescribed as
\begin{align}
	U_{in}(r,t) = \frac{(R-r)(R+r)}{R^2} U_{pulse}(t)  \qquad \text{with} \quad U_{pulse}(t) =
	\begin{cases}
		3.75 \, \left( 1-\cos\left( \frac{2 \pi }{0.003}t \right) \right) & \quad \text{if } t \le 0.003 \,,\\
		0 																						  & \quad \text{if } t > 0.003\,,
	\end{cases}
\end{align}
where $r$ is the radial coordinate direction.
As the plot of $U_{pulse}(t)$ in Figure \ref{fig:PressureTubeCase}b visualizes,
the inflow velocity is ramped up and down over a short period of time.

On account of its symmetry in circumferential direction,
the simulation is run on a $90^\circ$ section of the cylindrical domain only.
%
The fluid mesh consists of $14\,326$ linear Lagrangian tetrahedral elements with $3380$ nodes
and is refined close to the walls and the inlet.
The elastic channel wall is clamped at both ends and discretized by $6 \times 30 = 180$ nonlinear isogeometric Reissner-Mindlin shell elements \cite{Dornisch2013a} on a degree-2 NURBS
surface with $256$ control points.

The simulation is run for $300$ time steps of size $\Delta t = 2.5 \cdot 10^{-5}$. 
%
%
The convergence bounds are  again $\varepsilon_{Coupling}=10^{-6}$
and
$\varepsilon_{Problem}=10^{-10}$.

\subsubsection{Results}

The elastic tube case is
studied for two separate versions,
that differ in the boundary conditions:
(1) In the first configuration, the end of the pipe is closed by a rigid wall,
i.e., a fluid velocity of zero is prescribed.
Consequently, 
%
it again poses
a fully-enclosed problem
the Dirichlet-Neumann scheme cannot handle.
(2) The second case uses a free outflow condition at the pipe's end.
The results of both variants are illustrated in Figure \ref{Fig:TubeSnapshots} by three snapshots.
As expected, the increasing inflow velocity causes a widening of the tube close to the inlet.
When the velocity is reduced again,
this deformation starts to propagate like a wave through the elastic structure.
Qualitatively, a difference between the two configurations can mainly be observed towards the end of the simulation,
when the deformation pulse reaches the opposite end of the tube.

In the following, focus is again put on 
comparing 
the performance of different 
coupling schemes
for both setups.

\begin{figure}[h!]
	\centerline{\includegraphics[width=1.0\textwidth]{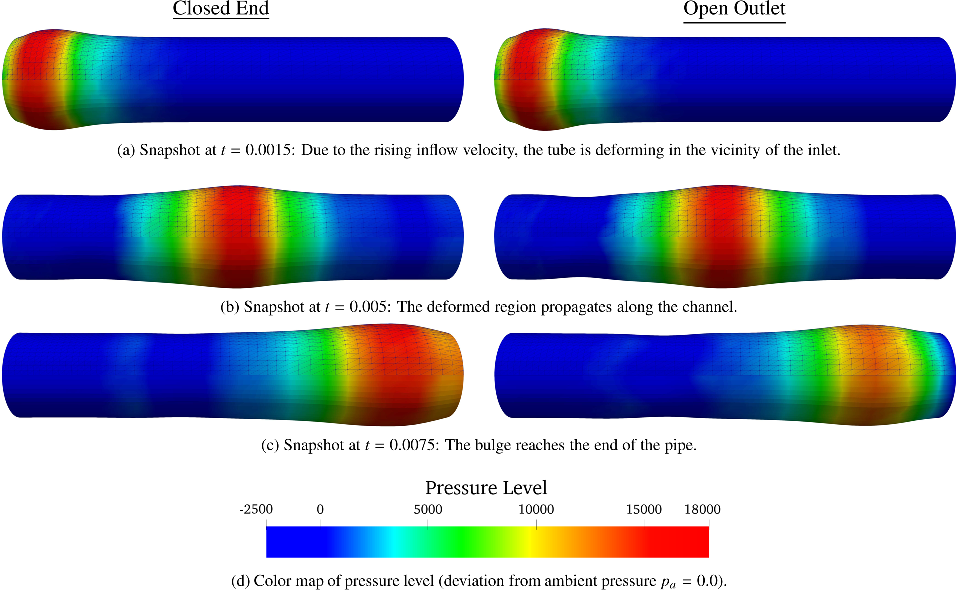}}	
	\caption[Elastic Tube: Propagating Shock]{Illustration of the deformation wave running through the tube for both test case versions.
		While the upper part of the pipe is the actual fluid mesh, the lower half follows from symmetry and is only shown for a better visualization.
		Coloring is based on the fluid's pressure field.}\label{Fig:TubeSnapshots}
\end{figure}

\subsubsection*{Closed End}

Looking at the results of various Robin-Neumann variants for the closed configuration 
in Table \ref{Tab:ClosedTube},
similar conclusions as for the previous test case can be drawn:
Again, 
the potential of the standard Robin-Neumann scheme
is strongly compromised by
its distinct sensitivity to the parameter $\alphaRN$.
This sensitivity is substantially 
reduced by the RN-QN scheme,
as it widely 
extends the parameter range that leads to
fast convergence.
Accordingly,
the efficiency gains are more significant for non-optimal values of $\alphaRN$, 
but the new RN-QN scheme clearly outperforms the plain Robin-Neumann approach for all choices.
With the IMVLS method,
it even provides faster convergence than the best result of the unmodified scheme
for almost the full parameter set.
It is worth noting that these major improvements come without any significant drawbacks.
Similar to the balloon test case, the accuracy of all Robin-Neumann variants is decreasing for lower values of $\alphaRN$,
independently from the quasi-Newton update.
However, the artificial flux is rather small for all simulations.

Again, an IQN method taking into account past time step data is the most effective choice.
Nevertheless, 
two alternatives are listed 
since their results 
point out the effects of poor choices for $\alphaRN$:
%
(1) If the parameter is too small, unnecessary weight is put on the Neumann term of the Robin condition, which slows down convergence.
%
As a secondary effect,
it also increases the number of coupling iterations required to construct
an accurate Jacobian approximation,
since subsequent data pairs are very similar.
The reutilization of past time step data is therefore particularly beneficial in this case.
Without it, the quasi-Newton update struggles
to a similar extent as Aitken's relaxation.
(2) Choosing $\alphaRN$ too high, 
on the other hand, 
increases the added-mass instability,
for which in general
quasi-Newton methods
are known to perform better than Aitken's relaxation \cite{spenke2020multi}.
Independent from the specific update technique, the Robin parameter 
shows an upper bound 
beyond which no stable results are obtained
due to the incompressibility dilemma,
as discussed in the next section.\\

In summary, the closed tube test case reinforces the 
key take-away of Section \ref{Sec:Ballon}, that is the superiority of the RN-QN scheme over the Robin-Neumann approach without update step.

\subsubsection*{Open Outlet}

So far, this work compared the new RN-QN scheme to the pure Robin-Neumann
coupling for two fully-enclosed cases.
For general FSI problems without fully-enclosed fluids, however,
the Dirichlet-Neumann scheme with interface quasi-Newton acceleration
is the prevalent partitioned algorithm.
Aside from multiple Robin-Neumann variants,
the performance analysis
of the tube with open outlet in Table \ref{Tab:OpenTube}
therefore
also lists
two Dirichlet-Neumann schemes
with quasi-Newton update:
As discussed in Remark \ref{Remark:DN_IQN_F}, the update is applied to either the interface deformation,
labeled as \textit{\dniqn{s}}, or the fluid forces, \textit{\dniqn{f}}.

First of all, looking only at the Robin-based approaches, the same advantages of the new RN-QN approach over the plain Robin-Neumann scheme as for the two previous test cases are observed.
A major difference, however, is that in this test case the RN-QN coupling does not have any apparent upper bound for the Robin parameter.
That is because 
as discussed in Section \ref{SubSec:RN},
increasing $\alphaRN$ brings the Robin-Neumann approach closer to the Dirichlet-Neumann scheme,
entailing its two main shortcomings:
Aside from the growing added-mass effect, which is kept under control by the quasi-Newton update,
the increased penalization of violating kinematic continuity essentially reintroduces the incompressibility dilemma.
As a consequence, the upper bound for the RN-QN scheme only exists in case of pure-Dirichlet problems,
since they in general cannot satisfy both
kinematic continuity and incompressibility at the same time.
For other FSI problems, like the tube with open outlet,
the incompressibility constraint will always be fulfilled,
so that kinematic continuity can be enforced by a very high Robin parameter without loosing stability.\\

The two Dirichlet-Neumann variants both yield good performance.
It is interesting to see that the \dniqn{f} algorithm converges slightly faster
than the more traditional \dniqn{s} variant.
An explanation could be that in this test case
it is a better initial guess 
to start a new time step with 
the same loads acting on the structural body,
rather than
%
the fluid experiencing the same boundary velocity
as in the previous time step\footnote{In the conducted simulations, the \dniqn{s} approach used a first-order extrapolation of the deformation as initial guess for the new time step.}.
%

The main focus of this test case, however, is to compare the new RN-QN scheme to the quasi-Newton-accelerated Dirichlet-Neumann approach:
It becomes apparent that
for the complete parameter range covered in the study,
the RN-QN coupling 
%
yields faster convergence.
This improvement is particularly pronounced for a range of about two orders of magnitude around $\alphaRN=10^{5}$,
which yields the biggest speed-up of roughly $50\%$.
In line with the interpretation of the Robin parameter, the convergence rate of the RN-QN scheme tends towards that of the \dniqn{f} strategy
for very high values of $\alphaRN$.
Despite the slight error in kinematic continuity inherent to Robin-Neumann variants,
these results of the RN-QN method
are remarkable,
considering
%
%
the widespread usage of
the Dirichlet-Neumann coupling with quasi-Newton update
in partitioned fluid-structure interaction.\\

\mypara 
Note that
even for the Dirichlet-Neumann scheme
the two solution fields will be consistent only up to a limited precision,
due to its staggered nature:
No matter when the iteration loop is left, one
solution field
was inevitably updated more recently than the other.
For the \dniqn{s} variant, the structural solver is called last, resulting in the negligible, though nonzero artificial flux seen in Table \ref{Tab:OpenTube}.
%
%
%
In contrast, the \dniqn{f} procedure exactly 
satisfies kinematic continuity, but has a marginal offset in the interface tractions.
%
Analogously, the RN-QN scheme's velocity difference is originating only from the Robin condition, not the staggered iteration.\\

\mypara 
In general, the Robin condition might have an effect on the convergence of the flow solver's internal Newton loop. Therefore, Table \ref{Tab:OpenTubeNewton} 
lists the average number of Newton iterations required per coupling step,
showing no significant influence of the Robin parameter.
Only the extremely high value of $\alphaRN=10^{18}$ yields a clear deterioration.



\section{Conclusion}

This work presents a novel coupling strategy for partitioned fluid-structure interaction
that merges 
the Robin-Neumann scheme with interface quasi-Newton methods.
%
%
As confirmed by the numerical examples in Section \ref{Sec:Results},
the \textit{Robin-Neumann quasi-Newton (RN-QN)} scheme benefits from the individual strengths of both its ingredients:
Being a Robin-Neumann approach, it can handle FSI simulations involving fully-enclosed incompressible fluids
and is less prone to the added-mass effect.
The interface quasi-Newton update, on the other hand, accelerates its convergence 
and widely extends the parameter range of $\alphaRN$ 
that produces stable results at all. 

The numerical studies further reveal one key advantage of the RN-QN scheme
that, in a sense, goes  beyond those of the original methods: 
It massively reduces the dependency on the Robin parameter
and therefore almost completely
overcomes the main drawback of the Robin-Neumann concept.
%
With that, finding a good choice for $\alphaRN$ may still be beneficial, but is no longer of primary importance.

Aside from that, the proposed direct feedback of fluid loads into the Robin condition
renders any explicit computation of structural Cauchy stresses obsolete
and hence
strongly facilitates the usage of Robin-Neumann schemes,
in particular for black-box solvers.

The comparison of different coupling algorithms in Section \ref{Sec:Results}
clearly shows 
the new RN-QN scheme is by all means preferable to the standard Robin-Neumann approach,
without introducing any drawbacks of its own.
Therefore, it is in particular beneficial for FSI simulations involving pure-Dirichlet fluid problems,
since in that case none of the Dirichlet-Neumann-based schemes provides an alternative.
Beyond that, the open tube test case allows to compare the new coupling algorithm to
the predominant partitioned approach for fluid-structure interaction, i.e., the
Dirichlet-Neumann scheme with an interface quasi-Newton update.
In a remarkable manner,
the RN-QN scheme not only keeps up 
for a very wide range of parameters,
but for good choices of $\alphaRN$ even outperforms it significantly.
These results 
demonstrate the potential of the new coupling to provide superior convergence speed. \\

In summary, the proposed \textit{Robin-Neumann quasi-Newton (RN-QN)} scheme is a new efficient and stable partitioned algorithm for fluid-structure interaction,
which in addition is capable of handling fully-enclosed incompressible fluids.


\section*{Acknowledgments}

The authors gratefully acknowledge the computing time granted by the JARA Vergabegremium and provided on the JARA Partition part of the supercomputer CLAIX at RWTH Aachen University.
This work is funded by the Federal Ministry of Education and Research (BMBF) and the state of North Rhine-Westphalia as part of the NHR Program.
%
%
%
%
%
%

%
%
%
%

\nocite{*}
\bibliography{references.bib}%


%


\begin{center}
\begin{table}[p!]
	\centering
	\caption{Parameters of the inflating balloon test case.} \label{Tab:BalloonParameters}
	\def\arraystretch{1.05}
	\begin{tabular}{ l  c  c c}
		\textbf{Parameter} & \textbf{Variable} & \textbf{Magnitude} & \textbf{Dimension} \\
		\hline
		Inner radius 				  & $r $		 & $0.05$ & [L] \\
		Initial outer radius 		& $R_0$		& $0.28$ & [L] \\
		\hline
		Fluid density				 &  $\rho_f$  &  $1000.0$ & [M L$^{-3}$] \\
		Dynamic fluid viscosity & $\mu_f$  &  $1.0$ & [M  L$^{-1}$ T$^{-1}$] \\
		Normal inflow velocity				& $U_{in}(t)$ & $10\sin(\pi t)$ & [M L$^{-1}$] \\
		\hline 
		Structural density &  $\rho_s$  &  $1000.0$ 					& [M L$^{-3}$] \\
		Young's modulus	& $E_s$			& $1.4 \cdot 10^6$	  & [M L$^{-1}$ T$^{-1}$]\\
		Poisson ratio		 & 	$\nu_s$		&  $0.3$					 & - \\
		Wall thickness				& $s$		   & $0.02$ & [L]
	\end{tabular}
\end{table}
\end{center}
 \begin{center}
	\begin{table}[p]%
			\caption[] {Inflating balloon test case: Comparison of the unmodified Robin-Neumann and the new RN-QN scheme. For RN-QN, the employed update scheme is indicated in brackets. The \textbf{bold numbers} indicate the average number of coupling iterations required per time step, the \textcolor{gray}{gray values} refer to the relative artificial flux $\varepsilon_{rel}$ as defined in Equation (\ref{Eqn:SigmaRel}). 
	Missing values indicate that a converged solution could not be obtained.} \label{Tab:BalloonResults}
%
%
	\bgroup
	\setlength{\tabcolsep}{6.2pt}
	\def\arraystretch{1.2}
	\begin{tabular}{  l l   c  c  c  c  c  c  c  c c c}
		\multicolumn{2}{l }{\multirow{2}{*}{\quad \textbf{Coupling scheme}}}	&	\multicolumn{10}{ c } {\textbf{Coupling iterations per time step} } \\ & &	\multicolumn{10}{  c } {\reldiff{\text{Relative artificial flux}}}  \\
		\hline 
		\parbox[][0.7cm]{1pt}{} \hspace{1.6cm} &  \multicolumn{2}{l}{$\alphaRN  =\quad ~10^3$} & $10^4$ &   $5 \cdot 10^4$ & $10^5$  & $5 \cdot 10^5$  & $10^6$ &  $10^7$ &  $10^8$  & $10^9$ &  $10^{10}$   \\ \cline{2-12}
		%
		\multicolumn{2}{l }{\multirow{2}{*}{\textbf{Robin-Neumann}}}  & 
		\iter{554.34} & \iter{73.05} & \iter{15.81} & 	\iter{6.33} &	\diverged	& \diverged &  \diverged			&  \diverged		&  \diverged	&  \diverged		\\
		& &	\reldiff{7.92\E{-4}}	&	\reldiff{1.22\E{-4}}	& 	\reldiff{2.53\E{-5}}	 &		\reldiff{1.30\E{-5}}		 &		&			&			&		&	    &  \\
		%
		\multicolumn{2}{l }{ \multirow{2}{*}{\textbf{RN-QN} (IMVLS)}}  & \iter{4.24}	& \iter{4.06}	& \iter{4.02}	&   \iter{4.00} & \iter{4.02} 	&  	\iter{4.03}		&		\iter{4.77}		& 	\iter{6.22}		& \iter{7.09}  &  \diverged		\\
& &	 \reldiff{7.92\E{-4}}	&	\reldiff{1.22\E{-4}}	& 	\reldiff{2.57\E{-5}}	 &		\reldiff{1.29\E{-5}}	& \reldiff{2.61\E{-6}}	&		\reldiff{1.30\E{-6}}		&	\reldiff{1.30\E{-7}}		&		\reldiff{1.30\E{-8}}	& 	\reldiff{1.30\E{-9}}	&    	\\
		%
		%
		\multicolumn{2}{l }{\multirow{2}{*}{\textbf{RN-QN} (ILS, q=0)}}  & \iter{17.49}	& \iter{8.09}	& \iter{5.24}	&  \iter{5.00}  & \iter{5.03} &	\iter{5.04} & \iter{5.11}	&  	\iter{7.80} & \iter{8.12} &  \diverged		\\	
& &	  \reldiff{7.92\E{-4}}	&	\reldiff{1.22\E{-4}}	& 	\reldiff{2.57\E{-5}}	 &		\reldiff{1.29\E{-5}}	& \reldiff{2.61\E{-6}}	&		\reldiff{1.30\E{-6}}		&	\reldiff{1.31\E{-7}}		&		\reldiff{1.30\E{-8}}	    & \reldiff{1.31\E{-9}} &	\\
		\multicolumn{2}{l }{\multirow{2}{*}{\textbf{RN-QN }(Aitken)}}  & \iter{44.83}	& \iter{11.45}	& \iter{7.02}	& \iter{6.00} &	 \iter{6.87} & \iter{6.97}	&  	\iter{7.12}		& \iter{10.18}	&	\iter{{16.04}}			&  \diverged				\\	
& &	  \reldiff{7.73\E{-4}}	&	\reldiff{1.22\E{-4}}	& 	\reldiff{2.57\E{-5}}	 &		\reldiff{1.30\E{-5}}	& \reldiff{2.61\E{-6}}	&		\reldiff{1.30\E{-6}}		&	\reldiff{1.31\E{-7}}		&		\reldiff{1.31\E{-8}} & \reldiff{1.30\E{-9}} & \\
	\end{tabular}
	%
	\egroup
\end{table}
\end{center}

 	\begin{table}[p]%
	\centering
	\caption{Closed tube test case: Comparison of various Robin-Neumann schemes. Again \textbf{bold numbers} are the average coupling iterations per time step, while the \textcolor{gray}{gray values} quantify accuracy via $\varepsilon_{rel}$ as defined in Equation (\ref{Eqn:SigmaRel}). Missing values indicate no stable results were found. \label{Tab:ClosedTube}}
	\setlength{\tabcolsep}{8pt}
	\def\arraystretch{1.2}
	\begin{tabular}{  l l   c  c  c  c  c  c  c  c c   }
	\multicolumn{2}{l }{\multirow{2}{*}{\quad \textbf{Coupling scheme}}}	&	\multicolumn{9}{ c } {\textbf{Coupling iterations per time step} } \\ 
	& &	\multicolumn{9}{  c } {   {\reldiff{\text{Relative artificial flux}} } } \\
	\hline 
	%
	\parbox[][0.7cm]{1pt}{} \hspace{1.6cm} &  \multicolumn{2}{l}{$\alphaRN  = \quad 10^3$} &  $10^4$ &  $5 \cdot 10^4$ & $10^5$  & $5 \cdot 10^5$ & $10^6$ &  $10^7$ &   $10^{8}$ &  $10^9$ \\ \cline{2-11}
	%
	\multicolumn{2}{l }{\multirow{2}{*}{\textbf{Robin-Neumann}}}  & 
	\iter{432.10}	&		 \iter{61.83} &	 \iter{12.30} & 	\iter{7.50} &  \diverged	& \diverged			&  \diverged &  \diverged			&	\diverged		\\
	& &	\reldiff{4.83\mathrm{E}{-3}}	&  \reldiff{ 8.84 \mathrm{E}{-4}} & \reldiff{ 2.18 \mathrm{E}{-4}}	 &	\reldiff{1.16\mathrm{E}{-4}}	&		&			&	&		&    \\
	%
	\multicolumn{2}{l }{ \multirow{2}{*}{\textbf{RN-QN} (IMVLS)}}  &
	\iter{6.58} & \iter{4.71}	& \iter{4.15}	& \iter{4.03} &	 	\iter{4.29}		&	\iter{4.57}	 & 	\iter{6.06}	 &  \iter{7.59} &  \diverged		\\
	&  & 	
	\reldiff{4.80\E{-3}}	&	\reldiff{8.67\E{-4}} &	 \reldiff{2.18\E{-4}}  &  \reldiff{1.16\E{-4}} 	&	\reldiff{2.51\E{-5}} 		&	\reldiff{1.27\E{-5}}	&  \reldiff{1.29\E{-6}}  	&	  \reldiff{1.29\E{-7}}
	 &  	\\
	%
	%
	\multicolumn{2}{l }{ \multirow{2}{*}{\textbf{RN-QN} (ILS, $q=0$)}}  &
	\iter{45.09} & \iter{17.14}	& \iter{8.05}	& \iter{6.95} &	 \iter{13.03} &	\iter{14.05}		&	\iter{15.49}	 & 	\iter{17.07}	 & \diverged	\\
	&  & 	
	\reldiff{4.80\E{-3}}	&	\reldiff{8.67\E{-4}} &	 \reldiff{2.18\E{-4}}  &  \reldiff{1.16\E{-4}} 	&	\reldiff{2.51\E{-5}} 		&	\reldiff{1.27\E{-5}}	&  \reldiff{1.29\E{-6}}  	&	  \reldiff{1.29\E{-7}}
	&   	\\	
	%
	\multicolumn{2}{l }{ \multirow{2}{*}{\textbf{RN-QN} (Aitken)}}  &
	\iter{56.22} & \iter{21.47}	& \iter{9.51}	& \iter{8.11} &	 	\iter{19.33}		&	\iter{26.02}	 & 	\iter{49.31}	 &  \iter{115.57} & \diverged \\
	&  & 	
	\reldiff{4.88\E{-3}}	&	\reldiff{8.67\E{-4}} &	 \reldiff{2.18\E{-4}}  &  \reldiff{1.16\E{-4}} 	&	\reldiff{2.51\E{-5}} 		&	\reldiff{1.27\E{-5}}	&  \reldiff{1.31\E{-6}}  	&	  \reldiff{1.30\E{-7}}
	&  	\\
\end{tabular}
\end{table}

	\begin{table}[p]%
		\centering
	\setlength{\tabcolsep}{7.2pt}
	\def\arraystretch{1.2}
	\caption{Open tube test case: Performance comparison of different partitioned algorithms, including both Robin-Neumann and Dirichlet-Neumann variants. (The coloring and formatting is equivalent to Table \ref{Tab:BalloonResults} and Table \ref{Tab:ClosedTube}.) } \label{Tab:OpenTube}
	\begin{tabular}{  l l   c  c  c  c  c  c  c  c c  }
		\multicolumn{2}{l }{\multirow{2}{*}{\quad \textbf{Coupling scheme}}}	&	\multicolumn{9}{ c } {\textbf{Coupling iterations per time step} } \\ & &	\multicolumn{9}{  c } { {\reldiff{\text{Relative artificial flux}} } } \\
		\hline 
		%
		\multicolumn{2}{l }{\multirow{2}{*}{\textbf{\dniqn{s}} (IMVLS)}}  & 	\multicolumn{9}{c}{ \iter{8.41}} \\
		& &	\multicolumn{9}{c}{ \reldiff{4.17\E{-12}}} \\
		%
		\multicolumn{2}{l }{\multirow{2}{*}{\textbf{\dniqn{f}} (IMVLS)}}  & 	\multicolumn{9}{c}{\iter{7.60}} \\
		& &	\multicolumn{9}{c}{ \reldiff{{0.0}}} \\ \cline{2-11}
		%
		\parbox[][0.7cm]{1pt}{} \hspace{1.6cm} &  \multicolumn{2}{l}{$\alphaRN   =  \quad 10^3$} &  $10^4$ &  $5 \cdot 10^4$ & $10^5$  & $5 \cdot 10^5$ & $10^6$ &  $10^7$ &   $10^{10}$ & $10^{18}$  \\ \cline{2-11}
		%
		\multicolumn{2}{l }{\multirow{2}{*}{\textbf{Robin-Neumann}}}  & 
		 \iter{432.49}		&		\iter{58.92} &	 \iter{11.60} & 	\iter{7.50} &  \diverged	& \diverged			&  \diverged &  \diverged			&	\diverged			\\
		& &	  \reldiff{5.03\E{-3}}	& \reldiff{9.28\E{-4}}	 	& \reldiff{ 2.39\mathrm{E}{-4}}	 &	\reldiff{1.28\mathrm{E}{-4}}	&		&			&	&		&    	\\
		%
		\multicolumn{2}{l }{ \multirow{2}{*}{\textbf{RN-QN} (IMVLS)}}  &
		 \iter{6.54} & \iter{4.70}	& \iter{4.14}	& \iter{4.07} &	 	\iter{4.44}		&	\iter{4.63}	  & \iter{5.81}	 &	\iter{6.78}  & \iter{7.20}	\\
		&  & 	
		\reldiff{4.84\E{-3}}	&	\reldiff{8.76\E{-4}} &	 \reldiff{2.31\E{-4}}  &  \reldiff{1.23\E{-4}} 	&	\reldiff{2.67\E{-5}} 		&	\reldiff{1.34\E{-5}}	&  \reldiff{1.36\E{-6}}  	&	
		 \reldiff{1.37\E{-8}} &  \reldiff{1.29\E{-12}} 	\\
	\end{tabular}
%
\end{table}

	\begin{table}[p]%
		\centering
	\caption{Open tube test case: Comparison of the Newton iterations per flow solver call. }\label{Tab:OpenTubeNewton}
	\setlength{\tabcolsep}{8pt}
	\def\arraystretch{1.5}
	\begin{tabular}{  l l   c  c  c  c  c  c  c  c c  }
		\multicolumn{2}{l }{\quad \textbf{Coupling scheme}}	&	\multicolumn{9}{ c } {\textbf{Newton iterations per flow solver call} } \\
		\hline 
		%
		\multicolumn{2}{l }{\textbf{\dniqn{s}} (IMVLS)}  & 	\multicolumn{9}{c}{ \iter{2.54}} \\
		%
		\multicolumn{2}{l }{\textbf{\dniqn{f}} (IMVLS)}  & 	\multicolumn{9}{c}{\iter{2.35}} \\ \cline{2-11}
		%
		\parbox[][0.6cm]{1pt}{} \hspace{1.6cm} &  \multicolumn{2}{l}{$\alphaRN   =  ~~~ 10^3~$} &  $10^4$ &  $5 \cdot 10^4$ & $10^5$  & $5 \cdot 10^5$ & $10^6$ &  $10^7$ &   $10^{10}$ & $10^{18}$  \\ \cline{2-11}
		%
		\multicolumn{2}{l }{{\textbf{Robin-Neumann}}}  &  \NewtonIter{2.12}	& \NewtonIter{2.31}	 	& \NewtonIter{2.57}	 &	\NewtonIter{2.63}	&	$\boldsymbol{-}$	&		$\boldsymbol{-}$	&	$\boldsymbol{-}$ &	$\boldsymbol{-}$	&    $\boldsymbol{-}$	\\
		\multicolumn{2}{l }{ {\textbf{RN-QN} (IMVLS)}}  &
		\NewtonIter{2.66}	& \NewtonIter{2.94} &	 \NewtonIter{2.70}  &  \NewtonIter{2.54}	&	\NewtonIter{2.92}	&	\NewtonIter{2.88}	&  \NewtonIter{2.77}	&	
		 \NewtonIter{2.71} &  \NewtonIter{5.25}	\\
	\end{tabular}
\end{table}

\end{document}